\documentclass[twocolumn]{aa}
\usepackage{graphicx}
\usepackage{txfonts}
\begin{document}
\title{The ubiquitous nature of the Horizontal Branch second $U$-jump}

\subtitle{A link with the Blue Hook scenario?
\thanks{Based on  observations with the ESO/MPI  2.2m and ESO/NTT
telescopes, located at La Silla Observatory (Chile) and on observations
with the NASA/ESA {\it Hubble Space Telescope}.}}

\author{Y. Momany\inst{1}, L. R. Bedin\inst{1}, S. Cassisi\inst{2}, 
G. Piotto\inst{1},  \\ S. Ortolani\inst{1}, A.   Recio-Blanco\inst{1},  F. De
Angeli\inst{1} \and F. Castelli\inst{3}}

\offprints{Y. Momany}

\institute {Dipartimento di Astronomia, Universit\`a di  Padova,   
Vicolo  dell'Osservatorio  2, I-35122     Padova, Italy\\
\email{momany,piotto,bedin,ortolani,recio,deangeli@pd.astro.it }
\and 
INAF - Osservatorio Astronomico di Collurania, Via M. Maggini,   
64100 Teramo,  Italy\\ 
\email{cassisi@te.astro.it}
\and
INAF - Osservatorio Astronomico di Trieste, via Tiepolo 11, 34131 
Trieste, Italy \\ \email{castelli@ts.astro.it}
}

\date{Received .., 2003 / accepted 19 February 2004}

\abstract{

In  a  previous paper we  reported  on a discontinuity in  the extreme
horizontal  branch (EHB)  of the  Galactic   globular cluster NGC6752,
which we called the second $U$-jump.  This feature was attributed to a
combination of post zero-age horizontal branch evolution and diffusion
effects.  In this follow-up study  we  analyze other EHB clusters  and
show that the  second $U$-jump is a  common feature among EHB clusters
reaching  $T_{\rm eff}\ge  23,000$K, and that   its onset in different
clusters converges around $T_{\rm eff}\sim 21,000\pm3,000$K.
We also present  near-ultraviolet diagrams of $\omega$Cen and NGC2808,
the only two   objects with spectroscopically confirmed  ``blue hook''
stars ($T_{\rm eff}\ge 35,000$K).
We confirm predictions of a photometric discontinuity
separating late  from   early-helium flashers.  Moreover,   we present
empirical evidence   that  the second  $U$-jump   population might be
mainly composed by early-helium flashers.
Lastly, we revisit the discussion on the ubiquitous nature of the gaps
and jumps  so  far  identified in  the blue   HB tails, suggesting   a
possible discrete nature of the  distribution in temperature of the HB
stars.

 }

\authorrunning{Momany et al.}

\titlerunning{The second $U$-jump}

\maketitle

\section{Introduction}

Over the last decades, both observational and theoretical efforts have
been devoted  to the analysis   of the observed  distribution of stars
along the Horizontal Branch (HB) of  Galactic globular clusters.  This
notwithstanding,  our understanding of  several observational features
is still  incomplete.  In fact,  even if theory and observations agree
that    the   HB morphology    is   governed  by    metallicity   (the
first-parameter),  since late sixties it  has  become clearer that the
color distribution of HB stars in Galactic globular  clusters is not a
unique function of metallicity (Sandage
\& Wildey \cite{sand67}). NGC362 and NGC288 are a classical example of  
how two clusters  sharing similar  metallicities can show   remarkably
different  HB morphology.  Hence,  other parameters (e.g. age, cluster
environment,    Helium abundance,  mass    loss and rotation)  besides
metallicity   affect  the evolution  of   HB   stars (second-parameter
debate).

\begin{table*}[t]
\begin{center}
\caption{Ground-based observations log along with the HST archival data}
\begin{tabular}{llllll}  
\hline\hline
\noalign{\smallskip}
Object & Instrument & Prog. ID & Date & Filters & Seeing \\
\hline
NGC1904 & WFI@2.2   & 64.L-0255 &  1999 & $U,B,V$   & 0\farcs8--1\farcs2 \\
NGC6752 & WFI@2.2   & 65.L-0561 &  2000 & $U,B,V$   & 0\farcs6--0\farcs9 \\
NGC7099 & WFI@2.2   & 65.L-0561 &  2000 & $U,B,V$   & 0\farcs6--0\farcs9 \\
NGC6273 & WFI@2.2   & 65.L-0561 &  2000 & $U,B,V$   & 0\farcs7--1\farcs1 \\
NGC7089 & WFI@2.2   & 69.D-0582 &  2002 & $U,B,V,I$ & 0\farcs7--1\farcs3 \\
NGC5139 & WFI@2.2   & 69.D-0582 &  2002 & $U,B,V,I$ & 0\farcs6--1\farcs5 \\
NGC5986 & SUSI2@NTT & 71.D-0175 &  2003 & $U,V$     & 0\farcs9--1\farcs4 \\
NGC6656 & SUSI2@NTT & 71.D-0175 &  2003 & $U,V$     & 0\farcs9--1\farcs4 \\
NGC6715 & SUSI2@NTT & 71.D-0175 &  2003 & $U,V$     & 0\farcs9--1\farcs4 \\
\hline
NGC2808                & WFPC2@HST & GO8655 & 2001 & $F450W$        &  \\ 
NGC2808                & WFPC2@HST & GO6804 & 1998 & $F336W$ &  \\ 
NGC6093		      & WFPC2@HST  & GO8655 & 2000 & $F450W$        & \\
NGC6093		      & WFPC2@HST  & GO6460 & 1997 & $F336W$        & \\
NGC6205		      & WFPC2@HST  & GO5903 & 1996 & $F336W$,$F450W$,$F555W$ & \\
\noalign{\smallskip}
\hline
\end{tabular} 
\label{t_log}
\end{center}
\end{table*}

The second parameter  effect is not the only  puzzling feature  in the
evolution  of HB stars.  In particular,  photometric  studies of stars
hotter than the RR Lyrae instability strip showed the presence of: (a)
gaps along the blue tail (Ferraro et al.\ \cite{ferr98}; Piotto et al.
\cite{piot99};  (b)  a  jump  around  $T_{\rm eff}\sim11,500$K in the
Str\"omgren $u$, $u-y$ (Grundahl et al.\  \cite{grun99}, hereafter G99)
and Johnson $U$, $U-V$ (Bedin et al.\ \cite{bedin00}) color-magnitude
diagrams (CMDs); (c) hot HB stars reaching temperatures of $T_{\rm
eff}\simeq30,000$K or more in metal-poor (D'Cruz et al
\cite{cruz96}, Brown et al.\  \cite{brow01}) and metal-rich (Rich
et al.\ \cite{rich97})   clusters;  and  (d) the   still   unexplained
presence of fast HB rotators (Behr et al.\ \cite{behr00}; Recio-Blanco
et  al.\  \cite{ale02}).   On the other   hand,  spectroscopic studies
showed  the presence  of  abundance (Behr et  al.\ \cite{behr99}), and
gravity anomalies (Moehler et al.\ \cite{moeh00}) in stars hotter than
$T_{\rm eff}\sim11,500$K.

Horizontal branch stars  hotter than   $T_{\rm eff}\sim20,000$ K   are
usually referred to  as extreme HB  (EHB) stars.  It is  believed that
EHBs experience high mass-loss  during their red giant phase, reducing
their H-rich envelope down to  $\le 0.05$M$_{\odot}$, to the point  of
being  unable to  sustain  H-shell burning.   However,  it  is hard to
explain  why such an enhanced  mass  loss occurs  along the red  giant
branch.
Near-UV  CMDs of NGC6752 (Momany  et al.\ \cite{moma02}) have revealed
another  interesting feature  along  the EHB.    In the  $U$ {\it vs.}
$(U-V)$  plane,  the   HB showed  a discontinuity   at $U-V\simeq-1.0$
(corresponding $T_{\rm  eff}\sim23,000$  K).  Given  the  (1) apparent
photometric similarities with  the  ``cooler'' G99 jump, and   (2) the
clear difference in temperature with respect  to blue hook stars (i.e.
blue hook stars are  generally hotter than $T_{\rm  eff}\sim35,000$ K)
we called    this feature  the ``second-$U$ jump'',    and tentatively
attributed it to  a combination of post  ZAHB evolution  and diffusion
effects.
%

Most of  these puzzling  features  remain unsolved.  In particular, we
lack of a global  view on the  origin and  internal properties of  EHB
stars. This is not a problem confined to the final stages of evolution
of globular cluster   stars.  Indeed, the  nature of  EHBs has a  more
general relevance in astrophysics as  these are considered responsible
of  the  UV excess observed    in the spectra  of elliptical  galaxies
($UV$-upturn galaxies, Greggio \& Renzini \cite{gregg90}).

In  this paper we  present new near-UV  CMDs  for a selected sample of
Galactic globular  clusters, characterized by  an HB with  an extended
blue tail, with the aim to investigate the observational properties of
the EHB stars in these clusters.  This  paper is organized as follows:
in the following section we  discuss  the observational data-base  and
briefly  outline the  main  reduction  and calibration  procedures; in
Section~\ref{s_diagrams} we show   that the  second $U$-jump,  already
identified   in  NGC6752, is also  present  in  other EHB clusters; in
Section~\ref{s_bh}   we  suggest a  link  between  the second $U$-jump
feature and the He flash induced mixing scenario discussed by Brown et
al.\ (\cite{brow01}).  A summary will close the paper.

\begin{table*}[t]
\begin{center}
\caption{Properties of cluster sample. Columns 2 to  5 are from the Harris 
on-line-catalog:    http://physun.physics.mcmaster.ca/~harris/mwgc.dat
(\cite{harris96}) as updated  on February  2003.  Columns 7  and 8 are
from Rosenberg et al.\ (\cite{rosen99})}
\begin{tabular}{llllllclll}  
\hline\hline
\noalign{\smallskip}
Object    & $E_{B-V}$ &  $(m-M)_V$  & [Fe/H] &   c$^{\mathrm{1}}$  & M$_V$ & $\Delta V_{TO}^{HB}$ & Normalized relative age & second $U$-jump & blue hook\\
\hline
NGC6752   & 0.04    & 13.13 & $-$1.56  & c                     &  -7.73   & 3.55 & 1.03 &     y  & n  \\ 
NGC6656   & 0.34    & 13.60 & $-$1.64  & 1.31                  &  -8.50   & 3.55 & 1.04 &     y  & n  \\ 
NGC5139   & 0.12    & 13.97 & $-$1.62$^{\mathrm{2}}$ & 1.61    &  -10.29  & ---  & ---  &     y  & y  \\ 
NGC6205   & 0.02    & 14.48 & $-$1.54  & 1.51                  &  -8.70   & 3.55 & 1.02 &   y & y$^{\mathrm{3}}$ \\ 
NGC7099   & 0.03    & 14.62 & $-$2.12  & c                     &  -7.43   & ---  & ---  &      n  & n  \\ 
NGC288    & 0.03    & 14.83 & $-$1.24  &  0.96                 &-6.74   & 3.55 & 0.97 & n & n \\
NGC7089   & 0.06    & 15.49 & $-$1.62  & 1.80                  &   -9.02  & ---  & ---  &     y  & n  \\
NGC6093   & 0.18    & 15.56 & $-$1.75  & 1.95                  &   -8.23  & 3.55 & 1.04 &     y  & n  \\ 
NGC2808   & 0.22    & 15.59 & $-$1.15  & 1.77                  &   -9.39  & 3.30 & 0.81 &     n  & y  \\ 
NGC1904   & 0.01    & 15.59 & $-$1.57  & 1.72                  &   -7.86  & 3.50 & 1.00 &     n  & n  \\
NGC6273   & 0.41$^{\mathrm{4}}$   & 15.95 & $-$1.68  & 1.53    &   -9.18  & ---  & ---  &      y  & n  \\ 
NGC5986   & 0.28    & 15.96 & $-$1.58  & 1.22                  &   -8.44  & ---  & ---  &      y  & n  \\ 
NGC6715   & 0.15    & 17.61 & $-$1.58$^{\mathrm{2}}$ & 1.84                  &   -10.01  & ---  & --- &      y  & y$^{\mathrm{3}}$  \\ 
\noalign{\smallskip}
\hline
\end{tabular} 
\label{t_clusters}
\end{center}
\begin{list}{}{}
\item[$^{\mathrm{1}}$] {Central concentration, c = log(r$_t$/r$_c$); a ``c'' denotes a core-collapsed cluster}
\item[$^{\mathrm{2}}$] {Refers to the metallicity of the dominant population}
\item[$^{\mathrm{3}}$] {We present evidence of the presence of blue hook stars}
\item[$^{\mathrm{4}}$] {A cluster suffering severe differential reddening}

\end{list}

\end{table*}


\section{Observations and Data Reductions}
\label{data_red}

In Table~\ref{t_log} we report the observation log of all data used in
this  paper,    while in  Table~\ref{t_clusters}  we  list    the main
properties, such as metallicity,  reddening, apparent distance moduli
and concentration, for the   selected clusters.  Here follows a  short
description of the data reduction process.

\subsection{Ground-based data}
\label{s_ground_log}

The ground-based data consist of $UBVI$ observations obtained in three
different  runs with the Wide-Field  Imager (WFI) at the 2.2~m ESO-MPI
telescope and one run using SUSI2 at the  ESO-NTT (both telescopes are
located in La Silla, Chile).

The WFI camera consists of   eight 2048$\times$4096 EEV-CCDs, with   a
total field  of view of $34\times33$   arcmin$^2$.  The exposure times
were divided in {\it deep}  and {\it shallow}  in order to sample both
bright red giants and the faint main sequence and HB  stars (e.g.  the
$U$ images of NGC2808 were  taken as a series  of 180 and 2000 second
exposures).  All scientific images were dithered in a way to cover the
gaps separating   the   eight 2048$\times$4096     CCDs.  The   seeing
conditions were generally good all over the three  runs (a typical run
consisted of   3     nights).  Shallow exposures     were obtained  in
photometric nights,  and these were  used to calibrate the  {\it deep}
data obtained  in conditions of thin  cirrus.  Basic reductions of the
CCD mosaic (de-biasing and  flat-fielding)  was performed using   the
IRAF package MSCRED (Valdes \cite{vald98}).

Stellar photometry   was  performed using   the  DAOPHOT and  ALLFRAME
programs (Stetson \cite{stet94}). For a detailed presentation we refer
the reader   to  Momany  et al.\    (\cite{moma02})  and Bedin   et al.
(\cite{bedin00}).  In order to perform stellar  photometry for 8 CCDs,
different  DAOPHOT/ALLFRAME  tasks   were included   in   routines  to
automatize the  reduction process.   This  pseudo photometric-pipeline
performed (1) the construction of  the point spread function (PSF) for
each single image; (2) initial fitting photometry;  (3) an estimate of
the aperture-correction from curves of  growth of the brightest stars;
(4) the construction of median image  for every chip, and the creation
of the master list of star-like objects;  and  (5) PSF fitting of
the master  list on  each single exposure.   The  {\it deep} and  {\it
shallow}   catalogs were then  matched,  and a  final  catalog in each
passband  was obtained.   To  reduce  the  identification of  spurious
objects, we imposed that objects recorded in the final catalog of each
passband  must  have been identified  in at  least half  of the single
exposures (i.e.  when merging 6  $V$ catalogs, only objects identified
in  at least 3 images were  registered).   The instrumental magnitudes
were  normalized to 1 second  exposure and zero airmass.  Finally, the
PSF magnitudes were converted  into aperture magnitudes assuming  that
$m_{\rm  ap}=m_{\rm PSF}-constant$, where  the   constant is the   the
aperture correction.

The photometric calibrations were defined  using standard $UBVI$ stars
from Landolt (\cite{land92}).   Secondary standard stars from  Stetson
(http://cadcwww.hia.nrc.ca/standards/) provided   a  larger  number of
standards. In a few cases these belonged to the same globular clusters
to  be calibrated.  However, Stetson's   fields lack $U$ measurements,
hence, in calibrating  $U$ data only  Landolt $U$  standards were
used.  Calibration  uncertainties in the  $UBVI$ filters are estimated
to be
%
%
$0.06$, $0.03$,  $0.03$ and  $0.04$  respectively.   In  deriving  the
calibration equations we assumed the following extinction coefficients
for   La   Silla:   $K_{U}=0.50$,    $K_{B}=0.23$,  $K_{V}=0.16$   and
$K_{I}=0.07$. It must be noted that the  small number of $U$ standards
makes the calibration of the $U$ data rather uncertain. In particular,
the possible presence of non-linear relations between instrumental and
standard magnitudes for very hot (blue stars) could not be checked.

The  NTT run (May 31,  2003) was specifically  designed to obtain high
quality UV diagrams in  order to check the  existence of the  second-$U$
jump in a number  of clusters.  At  the NTT we used  the SUperb-Seeing
Imager (SUSI2).   The  measured pixel scale  of  SUSI2 is $0\farcs085$
arcsec.  Each chip of the 2x1 mosaic  covers a field of 5.5$\times$2.7
arcmin$^2$.  The   seeing conditions were relatively  good (below
$1\farcs3$),   but unfortunately, this  night  was  non-photometric.  
From this  run, we  present excellent, but uncalibrated
CMDs for three  clusters (NGC5986, NGC6566 and  NGC6175), i.e. the NTT
data will  be   used only for comparative   purposes,  i.e.   {\it  no
quantitative descriptions will be drawn from these diagrams}.

%
%

\begin{figure*}
\centering
\includegraphics[width=16cm,height=16cm]{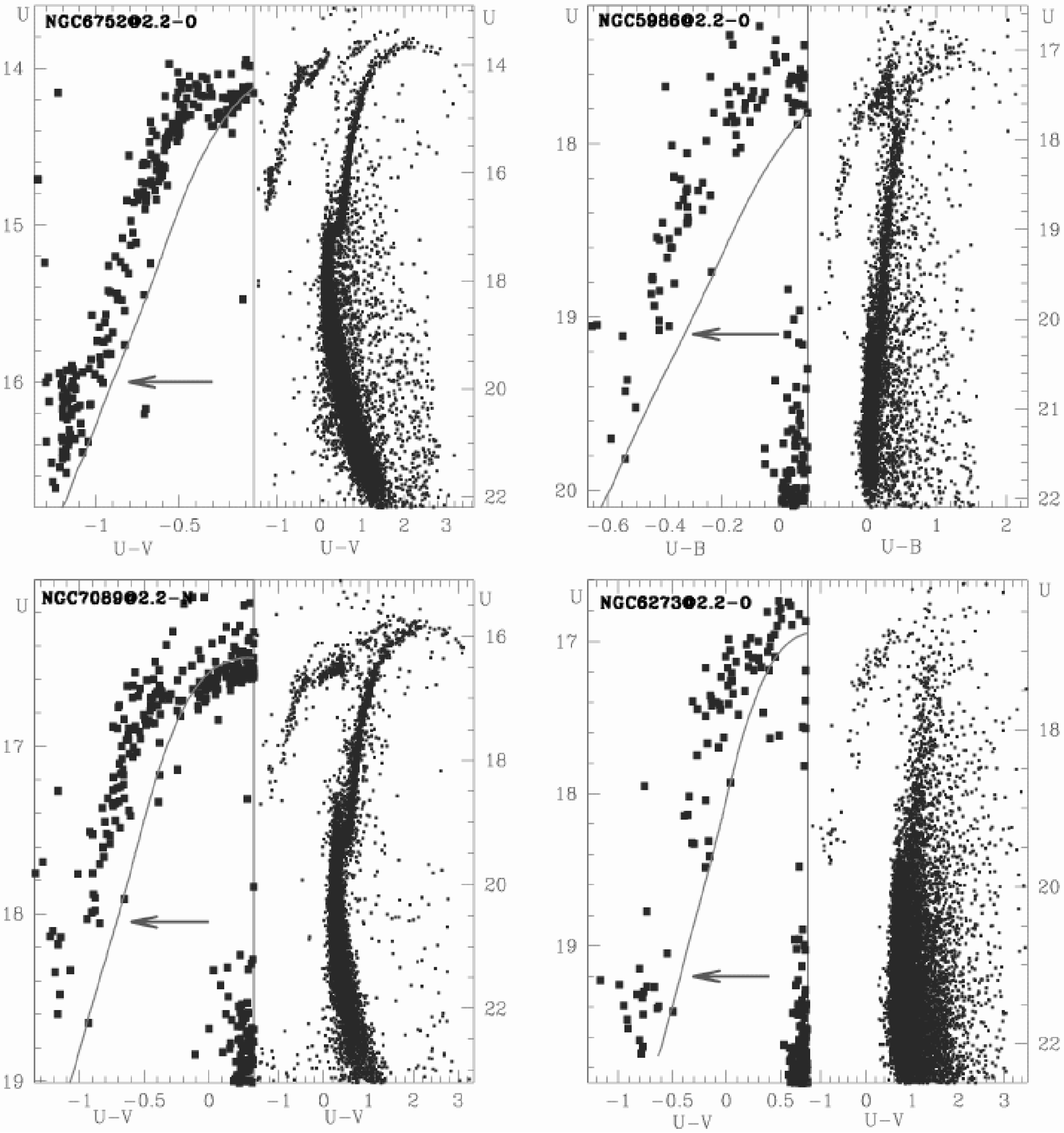}
\includegraphics[width=16cm,height=8cm]{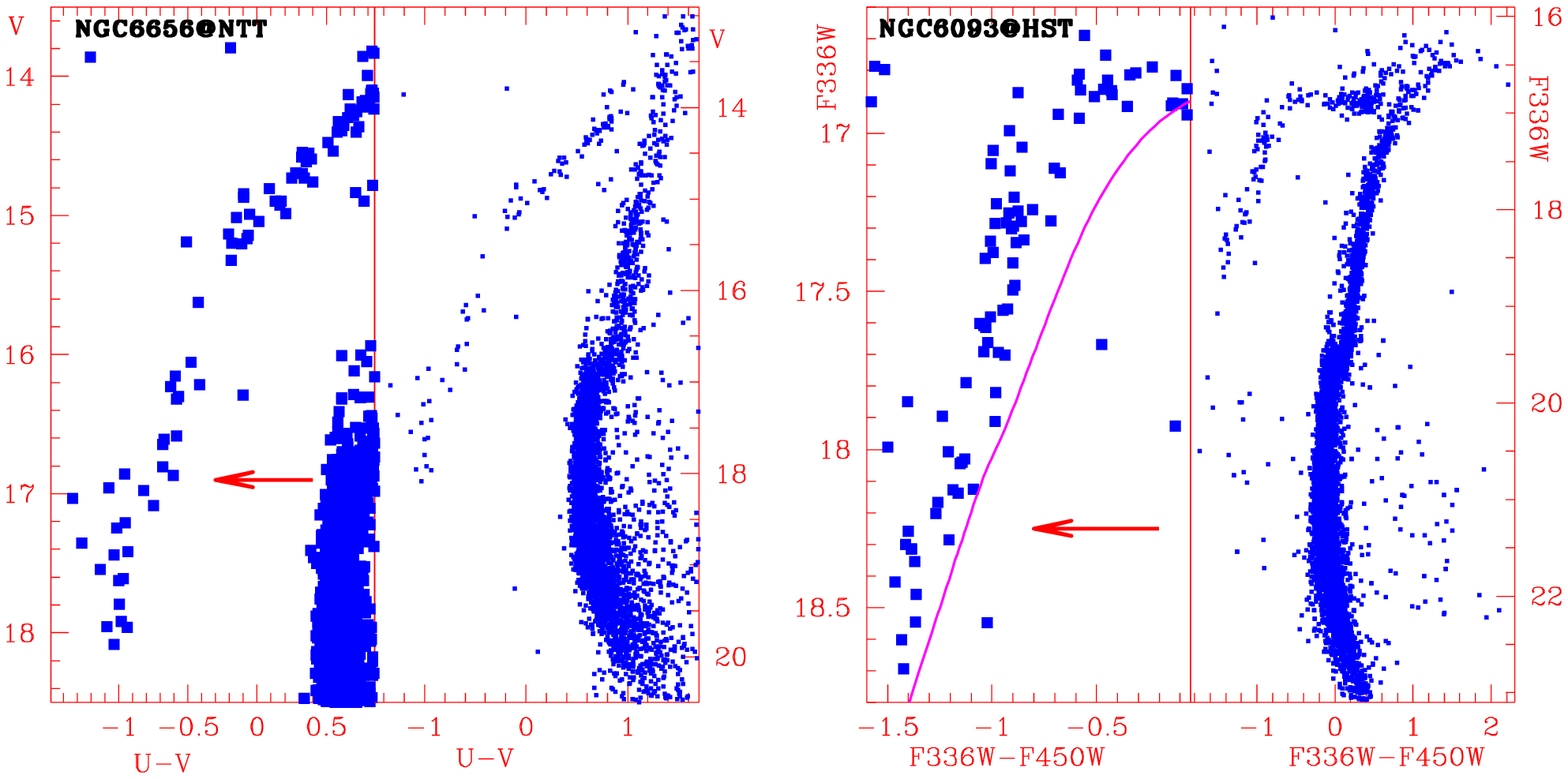}
\caption{Near ultraviolet diagrams of 6 EHB clusters.
The  right  panels show the  whole CMD   (highlighting  the high
completeness  level around the HB level),  while left panels show
a  zoom  on the HB  stellar  distribution, along with a $Z=0.006$
ZAHB model.    Arrows roughly   mark the  onset  of the  second
$U$-jump.}
\label{f_fig1}
\end{figure*}

Lastly, the  ground-based color-magnitude  diagrams presented  in this
paper    (e.g.  Fig.~\ref{f_fig1}), do  not   necessarily  show all the
collected photometric data.  Indeed, to avoid high crowding conditions
near cluster center and   outskirts  field contamination,    each
stellar catalog has been obtained by imposing  a selection on: (i) the
mean value  of  the image-shape statistics,  (SHARP,  normally between
$+0.5$ and $-0.5$); (ii) photometric  errors; and (ii) radius from the
cluster center.

\subsection{HST data}
\label{s_space_log}

We searched the  HST archive for $F336W$  and optical  observations of
all  EHB  globular clusters.    The importance  of multiple  dithering
(which allows a better derivation of the PSFs, Anderson \& King
\cite{ande00})  and non-adequate exposure times reduced the  number of  
useful clusters to only three, namely NGC2808, NGC6093 and NGC6205.

Following  the  methods     by  Anderson  \&  King     (\cite{ande00},
\cite{ande03}),   the photometric reduction  of  the $F336W$ and
$F450W$ observations has been carried out with algorithms based on the
effective point-spread-function (ePSF) fitting procedure.  The essence
of the method is  a finely sampled PSF of  high accuracy, created from
images having different offsets.  The   fit of the ePSF to  individual
star images  gives a precision in  position of the order of $\sim0.02$
pixel, with no systematic errors depending on the location of the star
with respect to the pixels boundaries. In principle a good positioning
implies good photometry  (not visa-versa), and this  can be  used as a
criteria to identify stars with good  photometry (as shown in Bedin et
al.\  \cite{bedin01}, \cite{bedin03}).   The  16 well-dithered $F450W$
images of     NGC2808  gave  us   an   ideal   case to     define  the
``good-positioning''  parameter    and served  best  our   purposes of
obtaining high  accuracy  PSFs.    This  allowed   us to  extract    a
``cleaned'' $F336W$, $F450W$ and $F555W$ catalog.

The instrumental (ePSF) magnitudes were converted to fixed aperture
photometry, and then calibrated to the WFPC2 synthetic system using
columns 2 and 3 from Table~9 in Holtzman et al.\ (\cite{holt95}).

\section{The near-UV Color-Magnitude Diagrams}
\label{s_diagrams}

In Fig.~\ref{f_fig1} we present near-UV CMDs  of 6 EHB  clusters,
namely NGC6752, NGC5986, NGC6273,  NGC7089, NGC6656, and NGC6093.   In
each panel we  report the origin of the  photometric data.   The right
panels  display the  global diagrams,  while left  panels show a
zoom on  the EHB  in order  to  highlight the  presence of  the second
$U$-jump.
For  data obtained at the 2.2m  telescope we pay attention to properly
identify the adopted filter  set (we use the  identifier ``O'' for the
old $U$ filter-set    and  ``N'' for  the  new  one):  Momany et   al.
(\cite{moma03}) have shown how  the  different transmission curves  of
the two  $U$ filters at  the 2.2m telescope translate into significant
differences on the slope, extension and morphology of blue HB stars.
Analyzing the CMDs of Fig.~\ref{f_fig1} we note the following:

\begin{itemize}

\item The global near-UV diagrams show clearly that we are not affected 
by incompleteness at the faintest level of the HB. The diagrams extend
down to $\sim 4-4.5$ mag below the main-sequence turn offs: i.e.  {\em
we reach the end of the EHB in all 
clusters};

\item All the CMDs are well populated in stars.
Wide-field imaging has surely helped in sampling more EHB stars. 
This  is  important  as we  intend   to study the    extension and the
morphology of the hottest, usually less populated part of the HB;

\item As indicated by the  arrows, 
there is a  discontinuity in the  EHB morphology (offset towards bluer
colors).  The HB   is displaced by   up to  $\sim   0.3$ mag in  color
(depending on  the employed color  and filters). Apparently, the bluer
displaced  part extends by   $\sim 0.5-0.7$  mag in  the $U$-magnitude
(again, depending on the employed filters);

\item In some clusters, the discontinuity is less evident due to 
the small number of stars, possible presence of differential reddening
and contamination  of  post-HB   stars. Differences in    the employed
photometric  systems  also contribute in  altering   the same features
along the HB (Momany et al.\ \cite{moma03});

\item It is worth noticing that the peculiar cluster NGC6273, 
known for its  high and differential  reddening, also shows the second
$U$-jump feature.   NGC6273 is also the  cluster with the largest
known  gap along  the HB  (see the optical   CMD  in the HST  snapshot
photometry by Piotto  et  al.~\cite{piot99}).  This brings about   the
discussion  of  whether there  is  a  connection between  the reported
discontinuities with the occurrence of gaps.
\end{itemize}

We argue that all clusters presented in Fig.~\ref{f_fig1} show evidence
of a discontinuity (second $U$-jump) in their  EHB, as that identified
in NGC6752. 

\section{An ubiquitous nature of the second U-jump ?}
\label{s_temperature}

In this section, we will verify whether the second $U$-jump is located
 at the same  physical position in all the  HBs.  If the onset of
the  second  $U$-jump is  observed  at  the same   location (e.g. same
temperature)  in more than   one     cluster, then one  can    exclude
statistical fluctuations as the cause of its appearance.
%
However,  before stating at what temperature  this occurs, we bring to
the  attention the difficulty in deriving ``photometric-temperatures''
along  the EHB.  In  Momany et  al.\  (\cite{moma02})  we estimated the
onset of the second $U$-jump  in NGC6752 ($U=15.90$ and $U-V=1.18$) at
$T_{\rm  eff} \sim  23,000$K.  This  estimate  was obtained  by simply
taking the  perpendicular projection of where  the onset of the second
$U$-jump is seen on an adequately shifted ZAHB model.  Considering the
models uncertainties and the larger bolometric corrections at the very
hot  end of the  EHB,  this  photometric-temperature  seemed close  to
spectroscopic determinations by  Moehler et  al.\ (\cite{moeh97}).   In
any  case, the uncertainty associated  to this temperature measurement
is of a few thousands degrees.


As  for  NGC6752, we derived photometric-temperatures  for  all of the
other clusters.  In order to  avoid other sources of uncertainties due
to  the distance modulus and reddening,  for each cluster we made sure
that the G99 jump  (clearly visible in  all  CMDs) was located  at the
expected temperature of $T_{\rm   eff} \sim 11,500$K.   The  resulting
temperature estimates   for the second  jump  are all in  the range of
$T_{\rm eff}=21,000\pm3,000$K.
Taking into account the uncertainties  associated to this measurement,
this is an indication that the onset of  the second jump is located at
the same temperature in all clusters of our sample.

Another indication  that the  second   jump is  located  at  the  same
temperature in {\em all} EHB clusters (presented in Fig.~\ref{f_fig1})
comes from the superposition of the clusters  HBs.  With the exception
of NGC2808, all our clusters   share the same metallicity within   0.2
dex, meaning  that the  metallicity  effects on  the HB morphology are
expected to be negligible.

The HB  superposition is  further facilitated  by  the occurrence of a
visible feature at a fixed  temperature, irrespective of the  examined
cluster:  the G99 jump at $\sim  11,500$K.  Although only qualitative,
this method showed  that the discontinuity in the  CMDs  occurs at the
same location, again suggesting    the ``ubiquitous'' nature   of  the
second jump  in all EHBs reaching  $T_{\rm  eff} \ge 22,000$K.   It is
important to note that we could not overplot  more than three clusters
at  a time.  This is  due  to  the fact  that   our cluster sample  is
observed at  three different  telescopes (NTT,  HST  and 2.2m)  with 4
different filter  sets.  Examples of  such clusters  superposition are
shown in Figs.~\ref{f_cen}, ~\ref{f_2808} and ~\ref{f_m54}.

\section{The second $U$-jump and the blue hook scenario}
\label{s_bh}

In discussing the occurrence of the second $U$-jump in NGC6752, Momany
et al.\ (\cite{moma02}) suggested that this feature  could be due to a
combination  of   post-ZAHB  evolution,   diffusion,  and    radiative
levitation  effects.  Having shown the  {\it ubiquitous} nature of the
second $U$-jump  in all clusters with EHB  exceeding $T_{\rm eff} \sim
22,000$K,   in  this section  we    present evidences  of  a  possible
connection   between  the second  $U$-jump  population   and the flash
induced-mixing scenario outlined by Brown et al.\ (\cite{brow01}).

In   the  past   few  years, a number of observations (D'Cruz  et al.
\cite{cruz00},  Whitney  et al.\ \cite{whit98}   and  
Brown et al.\ (\cite{brow01}) have  clearly shown that the hottest end
of the HB in $\omega$Cen and NGC2808  is populated by a peculiar class
of objects: after  Whitney et al.\  (\cite{whit98}) these  were called
blue hook stars.  A plain theoretical explanation for the evolutionary
origin of these stars - the so called He flash induced-mixing scenario
- has been provided by Brown et al.\ (\cite{brow01}). Up to date, this
is the most accredited theory on the nature of extremely hot HB stars,
and it has recently been supported by both theoretical (Cassisi et al.
\cite{cass03})     and     observational    efforts   (Moehler      et
al.\ \cite{moeh02}).

This scenario (see    also  Castellani \&  Castellani   \cite{cast93})
envisages  that,  as  a consequence  of  an  high mass-loss efficiency
during the red  giant branch evolution (due  to enhanced  stellar wind
and/or dynamical  interactions with other  stars  in the dense cluster
core), a  star can loose  so  much envelope  mass that  it fails to go
through Helium-Flash  at the tip  of the  giant branch, thus  evolving
toward the white dwarf  cooling  sequence with an  electron-degenerate
helium core.  Depending on the amount  of the residual H-rich envelope
mass, the star will ignite He-burning either
between the giant  branch tip and the   bright end of  the white dwarf
cooling  sequence ("early"  hot flasher),  or  along  the  white dwarf
cooling sequence ("late"  hot  flasher).   After the He-flash,   these
stars will settle on  the Zero Age   HB (ZAHB).  Given their  strongly
reduced envelope mass, they are much hotter than their counterparts on
the  canonical  ZAHB.   In addition,  in  a  late helium flasher
structure the convection  zone  produced by  the  late HEF is  able to
penetrate into the   H-rich envelope, thereby mixing  H  into  the hot
He-burning interior (He-flash mixing)  where it is  burned rapidly.  A
consequent dredge-up of processed material via  both H- and He-burning
enriches  the outer  envelope with  He and  some carbon  and nitrogen.
According   to  Brown et   al.\  (\cite{brow01}), these  abundance
anomalies cause    a  discontinuous increase    of the   HB  effective
temperature   at  the transition  between   unmixed  and mixed models,
producing  a  gap at the  hot  end of the  HB  stellar distribution as
indeed   observed    in  the  CMD   of   NGC2808   of   Bedin   et al.
(\cite{bedin00}).   At  the same   time, the changes   of  the surface
chemical  composition induced by the   He-flash mixing modify the
emergent  spectral energy distribution,  and may explain the fact that
these stars appear as sub-luminous objects in far-$UV$ CMD.


\subsection{The case of $\omega$Centauri}

\begin{figure*}[!ht]
\centering
\includegraphics[width=\textwidth]{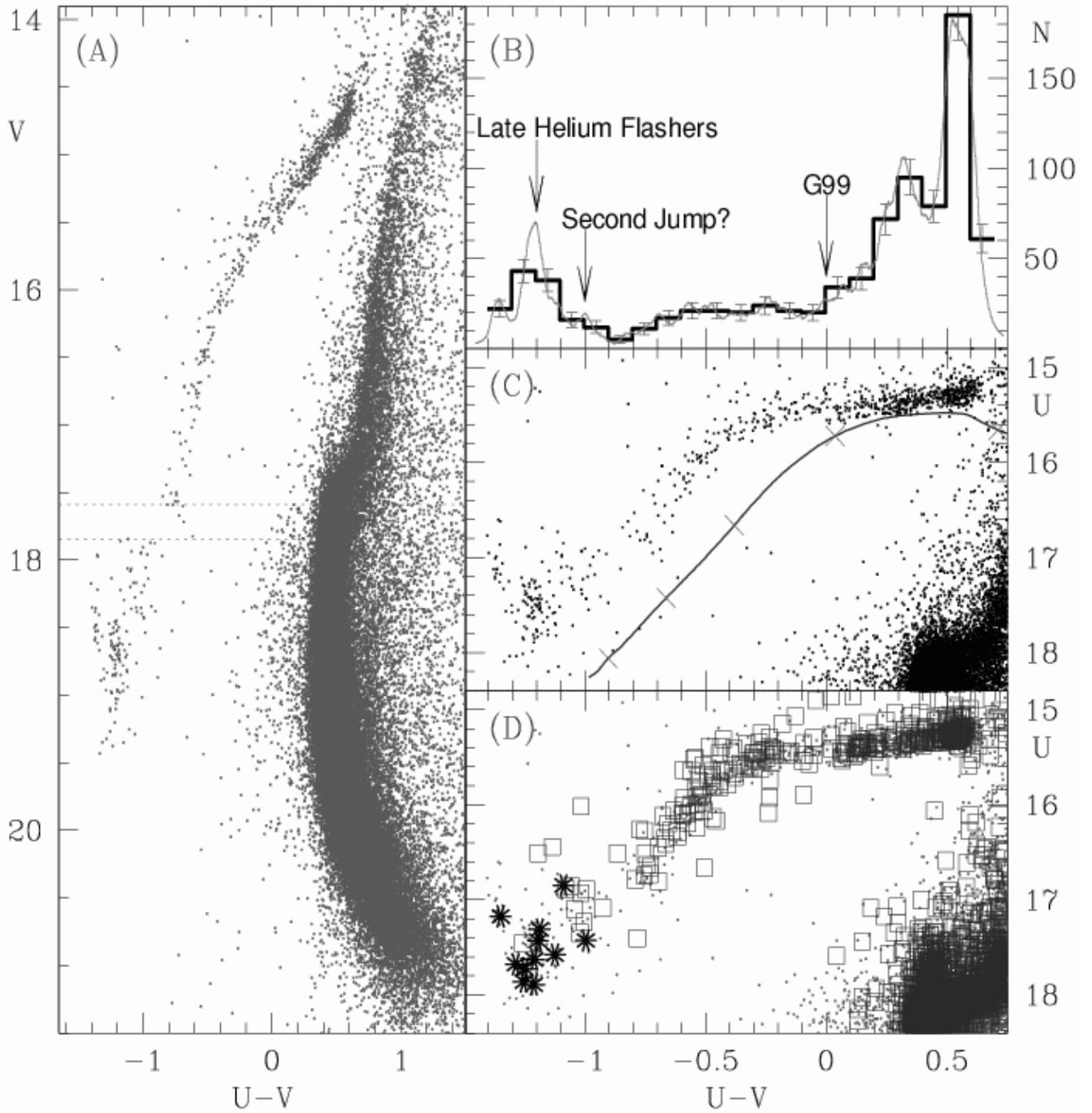}
\caption{Near-UV diagrams of $\omega$Cen: Panel (A) presents the global $V$, 
($U-V$) diagram.  The dashed  lines mark a  possible gap [see also the
$V$, ($V-I$) diagram in   Moehler et al.\ (\cite{moeh02})];  panel
(B) shows  the   ($U-V$)    color  distribution  histogram,    and   a
corresponding multi-bin, smoothed curve.    Panel (C) shows a  zoom on
the EHB along with  a  $Z=0.0006$ ZAHB  model.  Temperatures  of $6,000$,
$11,000$,  $16,000$, $23,000$ and $30,000$K  are marked  as crosses; and panel
(D) shows  an overplot  of the $U$,  ($U-V$)  diagram of NGC7089 (open
squares) on  that of  $\omega$Cen,    starred symbols  are  blue  hook
spectroscopic   stars     from     the  sample    of      Moehler   et
al.\ (\cite{moeh02}).}
\label{f_cen}
\end{figure*}

\begin{figure*}[!ht]
\centering
\includegraphics[width=\textwidth]{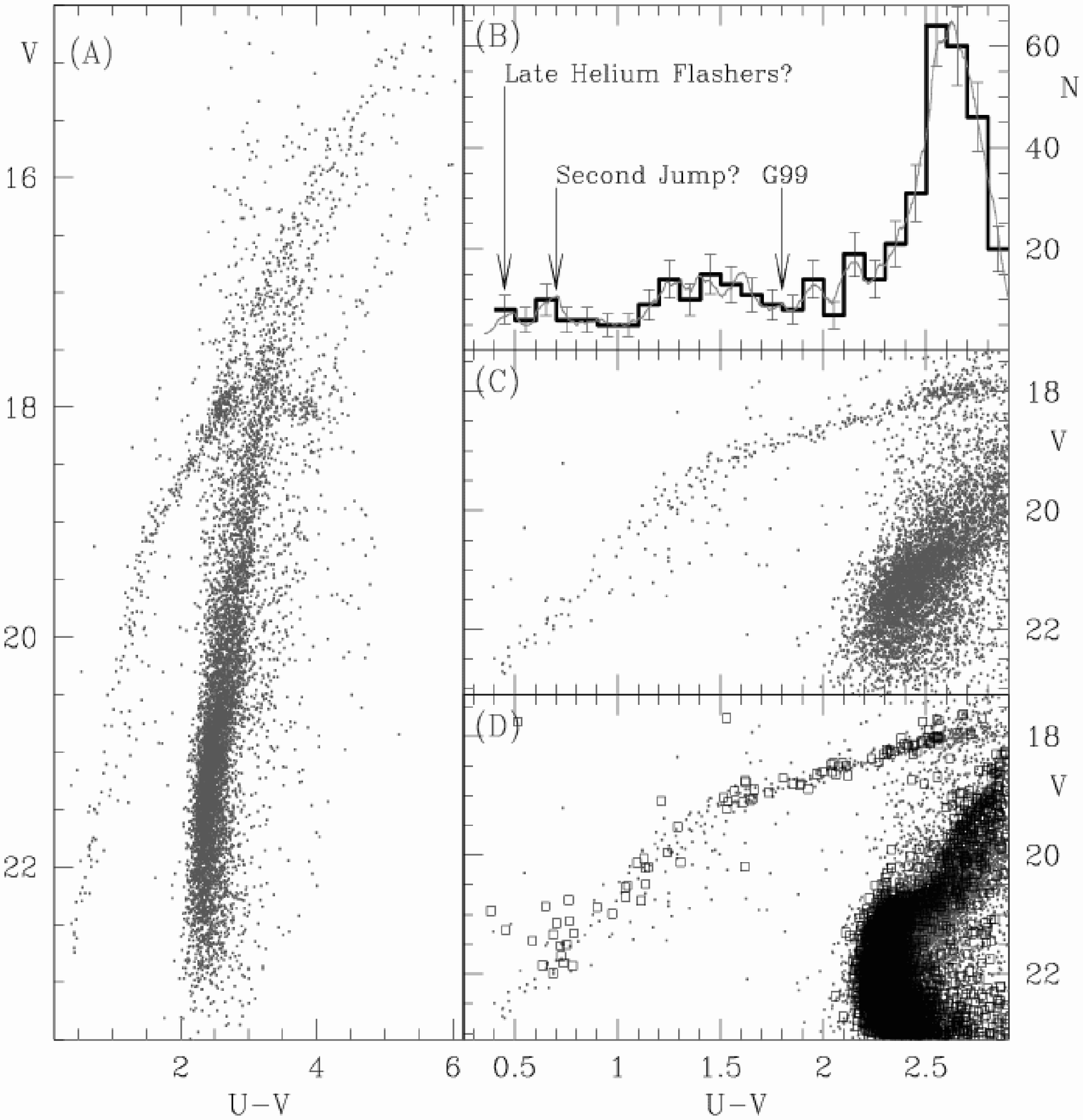}
\caption{NTT $V$, ($U-V$) diagrams of M54: (A) the global diagram; 
(B)  the  ($U-V$) color  distribution histogram,  and  a corresponding
multi-bin, smoothed curve; (C)  a zoom on  the extreme HB; and (D)  an
overplot of the $V$, ($U-V$) diagram of NGC6656 (open squares) on that
of M54. These are instrumental diagrams.}
\label{f_m54}
\end{figure*}

In Fig.~\ref{f_cen} we present  a high quality wide-field  near-UV CMD
of  $\omega$Cen.  Panel (A) shows a  blue HB that extends over $\sim5$
magnitudes in $V$.  In particular,  the HB extends $\sim1.6$ mag below
the turn off region, a similar behavior is seen in NGC2808 and NGC6715
(Figs.~\ref{f_2808} and ~\ref{f_m54}).
The region      between  $-0.8\le$($U-V$)$\le    -0.5$   is   sparsely
populated. Two horizontal (dotted) lines mark the only ``visible'' gap
seen     along   the   HB,     roughly    corresponding    to  $T_{\rm
eff}\sim19,000-20,000$K. This is the same gap seen in the $V$, ($V-I$)
CMD of Moehler et al.\ (\cite{moeh02}).  The end  of this gap is close
to the onset  of the ``second  jump'' population.   However, this does
not imply a direct gap/jump relation.  Indeed, the onset of the second
jump  in NGC6752 and NGC7089  is not preceded  by any visible gap.  In
panel (B) we  plot the  ($U-V$) color distribution  of  the HB.    The
continuous curve is the   smoothed, multi-bin,  color-distribution  as
done in Momany et al.\   (\cite{moma03b}).  The depression in  between
the  two peaks (centered at  surround ($U-V$)$=0.5$  and $0.7$) traces
the gap at  $10,000$K  probably hampered by photometric  errors, while
the  peak centered  at  ($U-V$)$=-1.2$ corresponds  to  the  blue hook
population, as suggested below.

A zoom  on the HB is shown in panels (C) and (D).
The HB in panel (C) seems to be divided into 5 different segments: (1)
{\it the blue HB segment}, which is the  part between the hot boundary
of the RR  Lyrae instability strip ($U-V\simeq  0.4$) and the G99 jump
($U-V\simeq 0.0$).  This segment  of  the HB  is well-reproduced  by a
canonical $Z=0.0006$ ZAHB model; (2)  the segment between $-0.8\le U-V
\le 0.0$.  Stars in this segment reach a maximum displacement from the
ZAHB model at $U-V\simeq-0.5$,  then  steadily re-approach the  model,
reaching it   around $T_{\rm eff}\sim20,000$K.    The same behavior is
seen in Fig.~\ref{f_fig1} by   other clusters; (3)  a group  of  stars
centered on $U-V\simeq-1.0$ and $17.8\le V \le 18.4$, this is probably
the second  $U$-jump  population; (4) a   group of stars   centered on
$U-V\simeq-1.2$ and $18.5\le   V \le 19.0$.    This is the  blue  hook
population previously identified   by Whitney et  al.\ (\cite{whit98})
and D'cruz et al.  (\cite{cruz00}) in far-UV studies.  Blue hook stars
are      characterized  by    their    high   temperatures    ($T_{\rm
eff}\ge35,000$K), thus they  appear  hotter/bluer than the ZAHB  model
stopping  at ($T_{\rm eff}\ge31,500$K);   and  lastly (5) a group   of
AGB-manqu\'e stars   obliquely  stretched  from  $U-V\simeq-1.25$  and
$V\simeq18.00$ to $U-V\simeq-1.35$ and $V\simeq16.70$.
%
%
%

\begin{figure}[!ht]
\includegraphics[width=9cm]{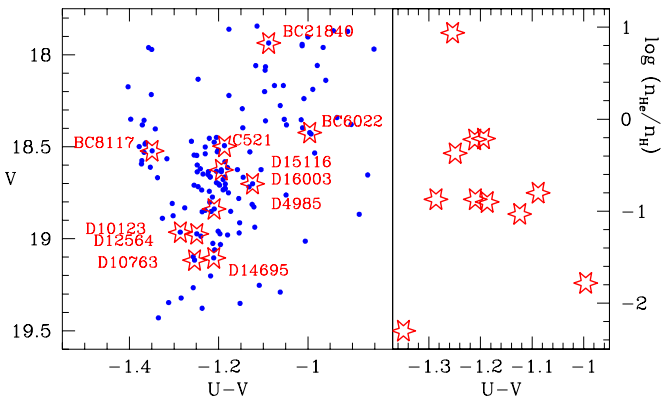}
\caption{Left panel shows the $V$, ($U-V$) diagrams of $\omega$Cen, 
along with  the blue  hook stars  (opens  stars) from  Moehler  et al.
(\cite{moeh02}).  Right panel plots the He  abundance (from Moehler et
al.\ \cite{moeh02}) $vs$ their ($U-V$) index.}
\label{f_moehler}
\end{figure}

Moehler et al.\ (\cite{moeh02})  listed a number of  $\omega$Cen stars
which have the  temperature and  chemical  signature expected for  the
blue hook stars.
%
\begin{table*}[t]
\begin{center}
\caption{Identification of the  Moehler et al.\  (\cite{moeh02}) spectroscopically confirmed  blue hook
stars  in  our $UBVI$ photometry  of  $\omega$Cen. Starred columns are
from  Moehler et al.\ (\cite{moeh02}).}
\begin{tabular}{llllllllll}  
\hline\hline
\noalign{\smallskip}
ID$^{*}$ & RA$_{(2000)}$ & Dec$_{(2000)}$ & $U$ & $B$ & $V$ & $I$ & $T_{eff}$$^{*}$  & log $g^{*}$ &    log ($n_{\rm He}$/$n_{\rm H}$)$^{*}$ \\
\hline
BC6022  & 13:26:24.936 & $-$47:33:23.42 & 17.427 & 18.365 & 18.423 & 18.620      & 45600$\pm$1300  & 6.10$\pm$0.14 & $-$1.78$\pm$0.16 \\
BC8117  & 13:26:33.215 & $-$47:35:12.56 & 17.173 & 18.322 & 18.522 & 18.776      & 29800$\pm$1000  & 5.48$\pm$0.14 & $-$2.30$\pm$0.23 \\
BC21840 & 13:27:25.945 & $-$47:32:19.79 & 16.848 & 17.788 & 17.936 & 18.205      & 35700$\pm$~~700 & 5.55$\pm$0.14 & $-$0.80$\pm$0.14 \\
C521    & 13:26:08.834 & $-$47:37:12.56 & 17.310 & 18.314 & 18.497 & 18.690      & 34700$\pm$~~500 & 5.90$\pm$0.12 & $-$0.90$\pm$0.09 \\
D4985   & 13:25:14.816 & $-$47:32:35.67 & 17.628 & 18.674 & 18.838 &  ---------  & 38400$\pm$~~800 & 6.08$\pm$0.16 & $-$0.87$\pm$0.16 \\
D10123  & 13:25:34.268 & $-$47:29:50.11 & 17.678 & 18.807 & 18.964 & 19.225      & 35000$\pm$~~500 & 5.82$\pm$0.12 & $-$0.87$\pm$0.09 \\
D10763  & 13:25:35.560 & $-$47:27:45.31 & 17.862 & 18.994 & 19.116 & 19.265      & 35200$\pm$1500  & 4.35$\pm$0.19 & $+$0.94$\pm$0.14 \\
D12564  & 13:25:41.316 & $-$47:29:06.30 & 17.724 & 18.820 & 18.973 & 19.179      & 36900$\pm$1000  & 5.60$\pm$0.14 & $-$0.37$\pm$0.09 \\
D14695  & 13:25:46.449 & $-$47:26:52.07 & 17.893 & 18.975 & 19.104 & 19.273      & 41300$\pm$~~700 & 6.11$\pm$0.21 & $-$0.22$\pm$0.12 \\
D15116  & 13:25:50.140 & $-$47:32:06.29 & 17.431 & 18.523 & 18.625 & 18.819      & 41500$\pm$1100  & 6.11$\pm$0.14 & $-$0.21$\pm$0.11 \\
D16003  & 13:25:53.953 & $-$47:35:21.64 & 17.577 & 18.604 & 18.701 & 18.878      & 36300$\pm$~~600 & 5.91$\pm$0.12 & $-$1.03$\pm$0.10 \\
\hline
\noalign{\smallskip}
\end{tabular} 
\label{t_Mohler}
\end{center}
\end{table*}
To  better understand  the  nature of  the  blue  hook stars,  we have
identified    in our photometry    of $\omega$Cen the   Moehler et al.
(\cite{moeh02})  stars.    Eleven   out   of  12 stars   were   found.
Table~\ref{t_clusters}  lists their  corresponding  magnitudes in  our
catalog  along with the effective  temperatures,  gravities and helium
abundances  from Moehler et   al.\ (\cite{moeh02}).  These objects are
labeled in  the left  panel  of Fig.~\ref{f_moehler}, while  the right
panel shows their  helium abundance plotted  against the ($U-V$) color
in our photometry.
Almost all the blue hook stars of the Moehler  et al.  (\cite{moeh02})
sample lie in the bluest and faintest  region of the $V$, ($U-V$) CMD.
There are however two exceptions: BC6022  and BC21840 are located near
the hot  end of  the canonical extreme   HB, showing  rather  ``cool''
($U-V$) indices  with respect to their  reported temperatures.  On the
other    hand, BC6022  seems   also  to   have  a   low helium content
($log(n_{\rm He}/n_{\rm H}=-1.78\pm0.16$), probably indicating a lower
flash mixing efficiency.  Finally, BC8117 has the bluest ($U-V$) color
in the  sample,  but a low temperature  ($T_{\rm eff}=29,800\pm1000$K)
and low He abundance  ($log(n_{\rm He}/n_{\rm H}=-2.30\pm 0.23$).   As
already  pointed out by Moehler et  al.\ (\cite{moeh02}), BC8117 could
be the descendant   of  an  early hot  flasher.    Clearly,  with  the
exception of BC8117, the helium content increases as the ($U-V$) index
decreases.
Therefore,  the identification   of the group  of   stars  centered at
($U-V$)$\simeq -1.2$ as the blue  hook population in $\omega$Cen seems
appropriate.  This also implies that the number  of blue hook stars in
$\omega$Cen is  far larger\footnote{There are  55 stars  centered at
$U-V\simeq-1.2$ with $18.5\le V \le 19.0$.  This is however is a lower
limit  for blue  hook  candidates  in  $\omega$Cen simply because  the
diagram we   present in Fig.~\ref{f_cen}   excludes the  central, most
crowded,  $5.5$   arcmin from our   $34 \times  33$ arcmin  catalog of
$\omega$Cen.}.
%
%
 Having clearly distinguished the blue hook population in $\omega$Cen,
 in panel (D) we overplot the CMD of NGC7089 (open squares) on that of
 $\omega$Cen.  Again, in  the matching process  we made sure  that the
 G99  jump of the two  clusters  coincided\footnote{The two  clusters
 were  observed  with   the  same  filter-set  at  the   ESO/MPI  2.2m
 telescope.}.  NGC7089 {\em does not possess}  a blue hook population,
 and panel (D) shows that its hottest stars,  {\em the second $U$-jump
 population}, coincide perfectly with the separated  group of stars in
 $\omega$Cen   that we    tentatively identify  as    the second  jump
 population.

In  order  to better understand    the relevance  of  the second  jump
population  in  the general picture of   extreme HB, and in  particular their
connection with the   He-flash mixing scenario,  one must bear in
mind the following facts:

\begin{itemize}

\item The second  $U$-jump population 
in NGC6752 extends  only  to the hot  end  predicted by  canonical  HB
models.    The same  applies    for the other   clusters  presented in
Fig.~\ref{f_fig1}.    None     of   the      clusters presented     in
Fig.~\ref{f_fig1}   exceeds  $T_{\rm  eff}\simeq31,500$K.         This
temperature is lower than the one predicted for blue hook stars in the
He-flash  mixing  scenario  ($T_{\rm eff}\ge36,000$K), hence   the two
features (blue hook and  second  jump populations) are quite  distinct
(see also below);

\item Second jump stars in NGC6752 are helium deficient 
(Moehler et al.\  \cite{moeh02});

\item The study of Moehler et al.\  (\cite{moeh02}) showed  that   blue hook   
stars   are indeed hotter  ($T_{\rm eff}\ge35,000$K) and are more helium-rich than 
classical canonical EHB stars.
A similar analysis  on the blue hook population  of NGC2808  even
showed carbon  enhancement in the  most  helium-rich stars (Moehler et
al.\  2003);

\item The flash-mixing scenario makes a clear distinction between early
and late hot flashers. Unlike the late flashers, early flashers manage
somehow to ignite helium  in between the giant  branch tip and the top
of the white  dwarf cooling sequence.  These also  manage to retain  a
strong enough  H-burning shell at  the point  of preventing any mixing
between  the helium   core  and hydrogen  envelope.  Consequently,  no
appreciable  changes occur  in their envelope   mass and  do  not show
altered chemical composition;

\item The flash-mixing scenario also predicts that  early and late hot
flashers  should     appear quite separated  in     the  CMD and, most
interestingly, having a sharp transition between the two populations.
The later feature,     a sharp transition between early  and   late
flashers, is exactly what is seen in our diagrams.  
Since  there is  a sizable population  of late  helium flasher in
$\omega$Cen, it  is reasonable (see also the  discussion in  Brown et
al.\  \cite{brow01})  to expect that a  significant population  of 
early helium flashers is also present. 
On theoretical grounds, early helium flashers are predicted to pile up
at  the  end of  the extreme  HB,  nearly  indistinguishable from {\em
canonical}  extreme HB stars.   Figure~\ref{f_cen} shows that the last
{\em  canonical} HB  stars and   prior  to the  onset of  late  helium
flashers (what we  call the second  $U$-jump population) {\em do} pile
up in a narrow color  range, exactly as late helium  flashers do.   We
therefore  argue  that a significant fraction   of the second $U$-jump
stars in $\omega$Cen are {\em bona-fide} early helium flasher objects.

\end{itemize}

In summary, on the basis of  the evidences quoted  above, there is the
possibility that the  second  jump stellar population in   $\omega$Cen
contains a significant number of  early helium flasher stars.  Whereas
it is  possible  to  investigate  the nature  of  late  helium flasher
objects  by   spectroscopical  measurements   (see   Moehler et   al.\
\cite{moeh02}), there is no way to directly discriminate between early
helium flashers  and {\it canonical} EHB  stars, since  no differences
are expected in their envelope chemical composition.
However, it is plausible  that the physical  mechanism able to enhance
the mass loss, producing both early and late helium flashers should be
the same.  One possibility is that mass loss in a giant evolving along
the red giant branch is enhanced by  the presence of a companion, i.e.
in  a  binary  system.   It would   be worthwhile  to  observationally
investigate for the occurrence of binarity  among both blue hook stars
and second jump  stars    in $\omega$Cen,  and eventually study    the
properties of these binaries.

\subsection{The case of NGC2808}

\begin{figure*}[!ht]
\centering
\includegraphics[width=\textwidth]{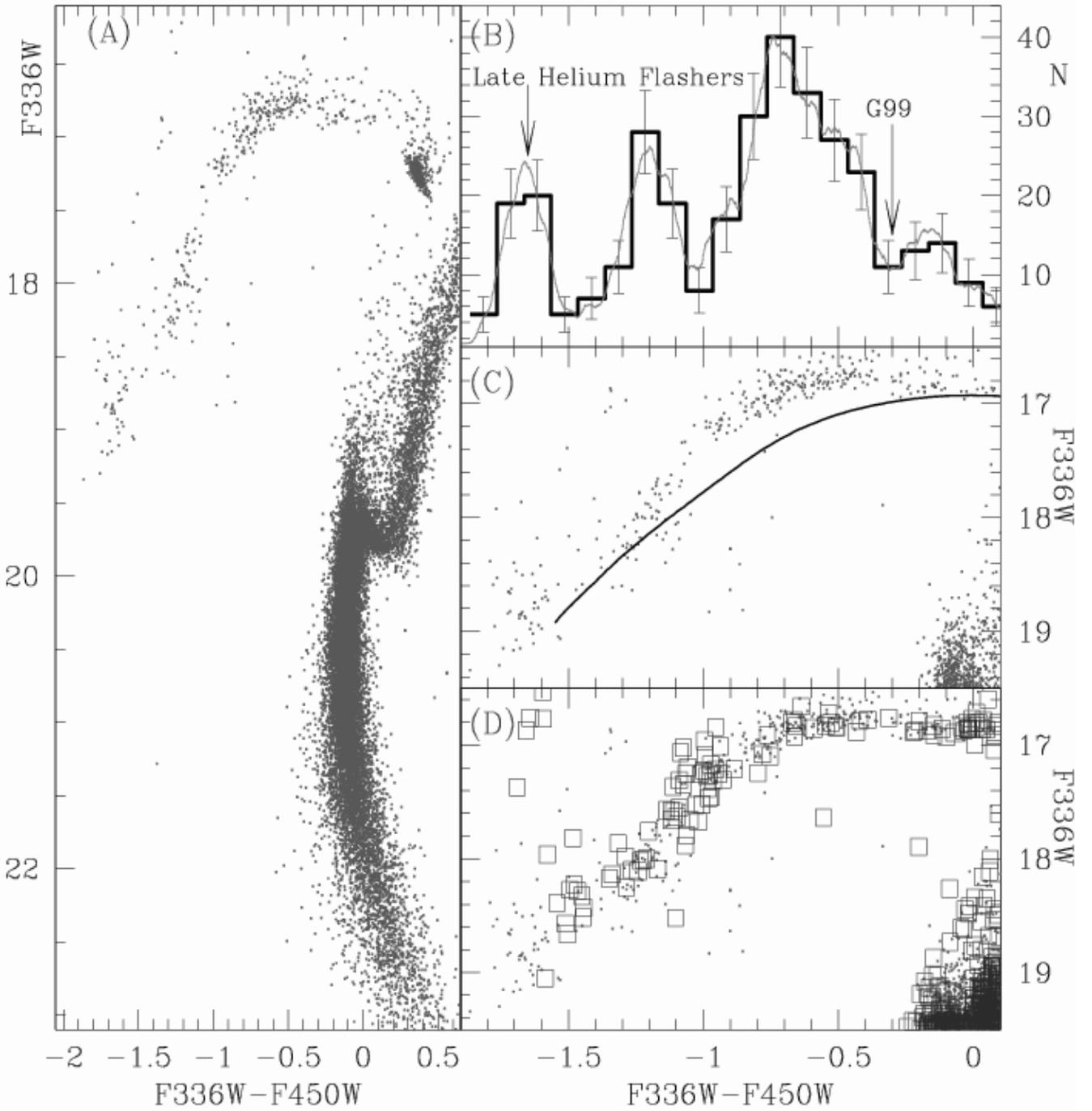}
\caption {HST $F336W$, ($F336W-F450W$) diagrams of NGC2808: 
Panel (A) presents the global $F336W$, ($F336W-F450W$) diagram.  Panel
(B) shows  the  ($F336W-F450W$) color  distribution histogram,  and  a
corresponding multi-bin,  smoothed curve.  Panel (C)  shows  a zoom on
the EHB  along with  a $Z=0.001$ ZAHB  model; and  panel (D)  shows an
overplot of  the $F336W$,   ($F336W-F450W$) diagram of   NGC6093 (open
squares) on that of NGC2808.}
\label{f_2808}
\end{figure*}

We  have  investigated   whether  $\omega$Cen  is   the only   cluster
possessing {\em both} early and late helium  flasher stars.  The first
candidate   to verify  this  hypothesis  is  NGC2808,  the  only other
globular  cluster with    a  spectroscopically  confirmed   blue  hook
population (see Moehler et  al.\  \cite{moeh03}).  Figure~\ref{f_2808}
presents HST $F336W$, ($F336W-F450W$) diagrams of NGC2808, done in the
same manner as in Fig.~\ref{f_cen}.
With respect to the far-UV diagrams  of Brown et al.\ (\cite{brow01}),
panels (A)  and  (B) show a   complete  sampling of both  hottest  and
reddest HB stars.
Indeed, one  can  easily detect  the   (1)  RR  instability strip
centered   at  ($F336W-F450W$)$\simeq0.2$;    (2)   G99  jump  at
($F336W-F450W$)$\simeq-0.35$;        (3)   gap      centered   at
$F336W\simeq17.4$  and   ($F336W-F450W$)$\simeq-1.15$  (Sosin   et al.
\cite{sosin97}), roughly  corresponding to  $T_{\rm  eff}\sim16,000$K;
(4) second gap located  between $-1.65\le$ ($F336W-F450W$)$\le -1.30$;
(5) late helium flashers (Brown et al.\ \cite{brow01}) centered around
($F336W-F450W$)$\simeq-1.65$, whose   temperature clearly exceeds that
of a   canonical  ZAHB model\footnote{Evolutionary  models  adopted in
present work have been computed  in the framework of canonical stellar
evolution, i.e.  neglecting  any non-canonical physical process.}   as
seen in  panel (C); and  lastly (6) group  of post-HB  stars extending
between $16.20 \le F336W \le 18.00$ at ($F336W-F450W$)$\le-1.3$.

Comparing   Figs.~\ref{f_cen}  and   \ref{f_2808}   one  notes   clear
differences   in  the two  color  distributions  (panels  B) and, most
interestingly, finds it hard to identify the second jump population in
NGC2808, which should occur at  the level of  the second gap.  This is
best seen in panel (D) where we overplot the CMD of NGC6093 on that of
NGC2808. Again the two CMDs have been matched  at the level of the G99
jump.    Panel (D) confirms  that   the second $U$-jump population  of
NGC6093 overlaps the gap in the extreme HB of NGC2808.
Though our  HB  models exclude  that  the  absence   of the second
$U$-jump population in  NGC2808 can  be  due  to   a metallicity
effect which changes the position of  the stars along  the HB, it must
nevertheless be   noticed that  NGC2808  is  the only  cluster  with a
significantly different metal   content   with respect to    the other
clusters in our sample.
The immediate conclusion of  this comparison is that  NGC2808, besides
being the {\em only  cluster} showing a  gap in its  extreme HB, it is
also  the  only  exception   to  the general  evidence   presented  in
Section~\ref{s_temperature}  that  {\em  all}   extreme  HB   clusters
reaching $T_{\rm eff}\ge22,000$K show the second $U$-jump population.

Assuming the case of $\omega$Cen (showing {\em both} a second jump and
late helium flasher populations) as a  {\it complete} manifestation of
the   flash-mixed scenario, then one  must  ask  why  NGC2808 seems to
literally {\it skip} the formation of  the second jump or early helium
flasher  population.   This   questioning   does   not conflict   with
interpretation given by  Brown et al.\ (\cite{brow01}),  who attribute
the gap within the extreme HB of NGC2808  to the dichotomy between the
blue hook and canonical extreme HB models.  So far, we lack for a full
explanation of this empirical evidence.
However, it is  worth mentioning that the  absence of  the second jump
population (or the presence  of the second gap) in  NGC2808 is not the
only   peculiarity  of  this  cluster.    In   fact,   despite  of the
considerable number  of both red and blue  HB stars (see panel B), and
the relatively high mass, NGC2808  is known to  possess {\em only two}
RR Lyrae stars   (Clement \&  Hazen \cite{clement89}).   Indeed,   the
reported ``Specific frequency of RR Lyrae variables'' in NGC2808 (0.3)
is among the lowest (Harris  \cite{harris96}).  The only other cluster
showing both red and blue HB  stars and such  a low RR Lyrae frequency
is NGC6441.

%

%
%

\subsection{The cases of M54 and NGC6205}

To further support the suggestion  that NGC2808 is  an exception to  a
scenario   in   which all   extreme   HB clusters   (reaching  $T_{\rm
eff}\ge23,000$K) should develop a  second jump population, we bring to
the attention the case of M54 (NGC6715).  This cluster  is in the core
of the Sagittarius dwarf galaxy (Ibata et al.
\cite{ibata97}).  Figure~\ref{f_m54}     presents    our NTT   near-UV
($\sim5.5\times5.5$ arcmin$^2$) diagram  of M54.  As mentioned  in
section~\ref{s_ground_log}, the calibration  uncertainties in  the NTT
data (due to a non-photometric night) should not prevent a comparative
analysis of the instrumental CMDs.

On the one hand, Panel  (A) shows that  the extension  of the blue  HB
below the turn off  region is very similar to  that of $\omega$Cen and
NGC2808;    i.e.  $\simeq1.6$ magnitude  in   $V$.   This  simple fact
suggests the presence   of    a late helium flasher    population   in
M54\footnote{A  similar conclusion has   been reached by  Rosenberg et
al.\ (2003), based  on $BV$ photometry.}. The  main point we here want
to emphasize is shown  in panel (D).  Overplotting  the CMD of NGC6656
(a cluster possessing  a  second jump  population but no   late helium
flashers) on M54,  one  sees that   the  location of the second   jump
population in NGC6656 overlaps with a group of stars in M54; i.e.  M54
seems  to   possess  both the  second jump    and late  helium flasher
populations.

Within the scenario  we proposed in  Section \ref{s_bh}, the HB of
$\omega$Cen and  M54 show evidence of  possessing {\em both} early  and late
helium flashers, whereas NGC2808 shows only  late flashers.  Obviously
with such a small   sample  of stellar  systems hosting  early and
late helium flashers   we cannot derive   sound conclusions regarding
when  the production of {\em both}  early  and late helium flashers is
the general rule or if it is an exception.

\begin{figure}[!ht]
\includegraphics[width=9cm]{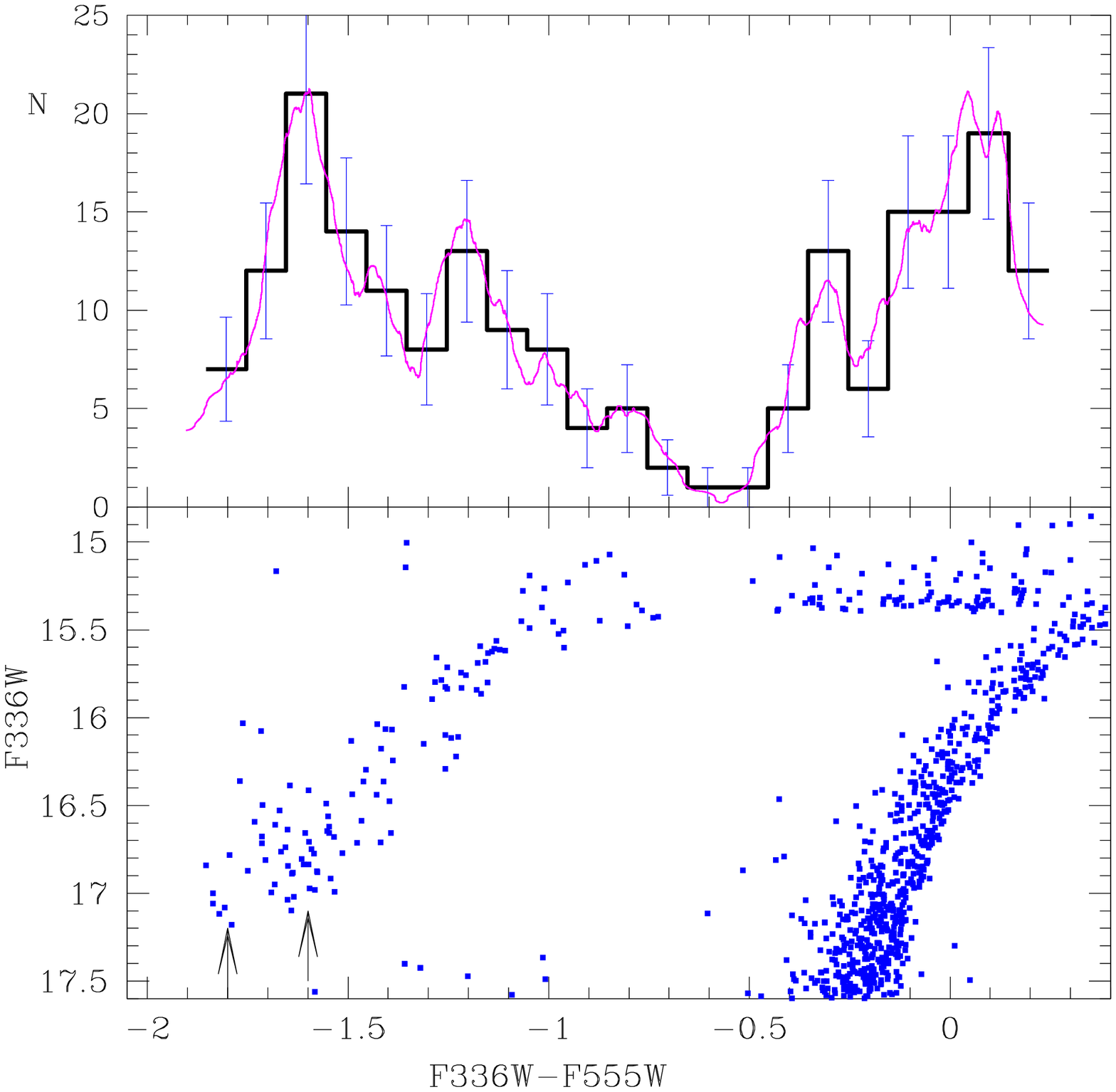}
\caption{HST $F336W$, $(F336W-F555W)$ diagram of NGC6205. Arrows 
mark the second jump population  and  a group of  hotter
stars.}
\label{f_6205}
\end{figure}

Lastly, we investigated also the possible  presence of blue hook stars
in NGC6205.  Figure~\ref{f_6205} presents HST $F336W$, $(F336W-F555W)$
diagram  of  NGC6205.   Arrows mark    the onset of   the second  jump
population  (see  also the  diagrams in  Grundahl et  al.  1999) and a
group of stars, offsetted towards {\em bluer} colors and {\em fainter}
magnitudes.  These could be AGB-manqu\'e stars as  well as late helium
flashers, and clearly call for a spectroscopic follow-up.

\begin{figure*}
\centering
\includegraphics[width=\textwidth]{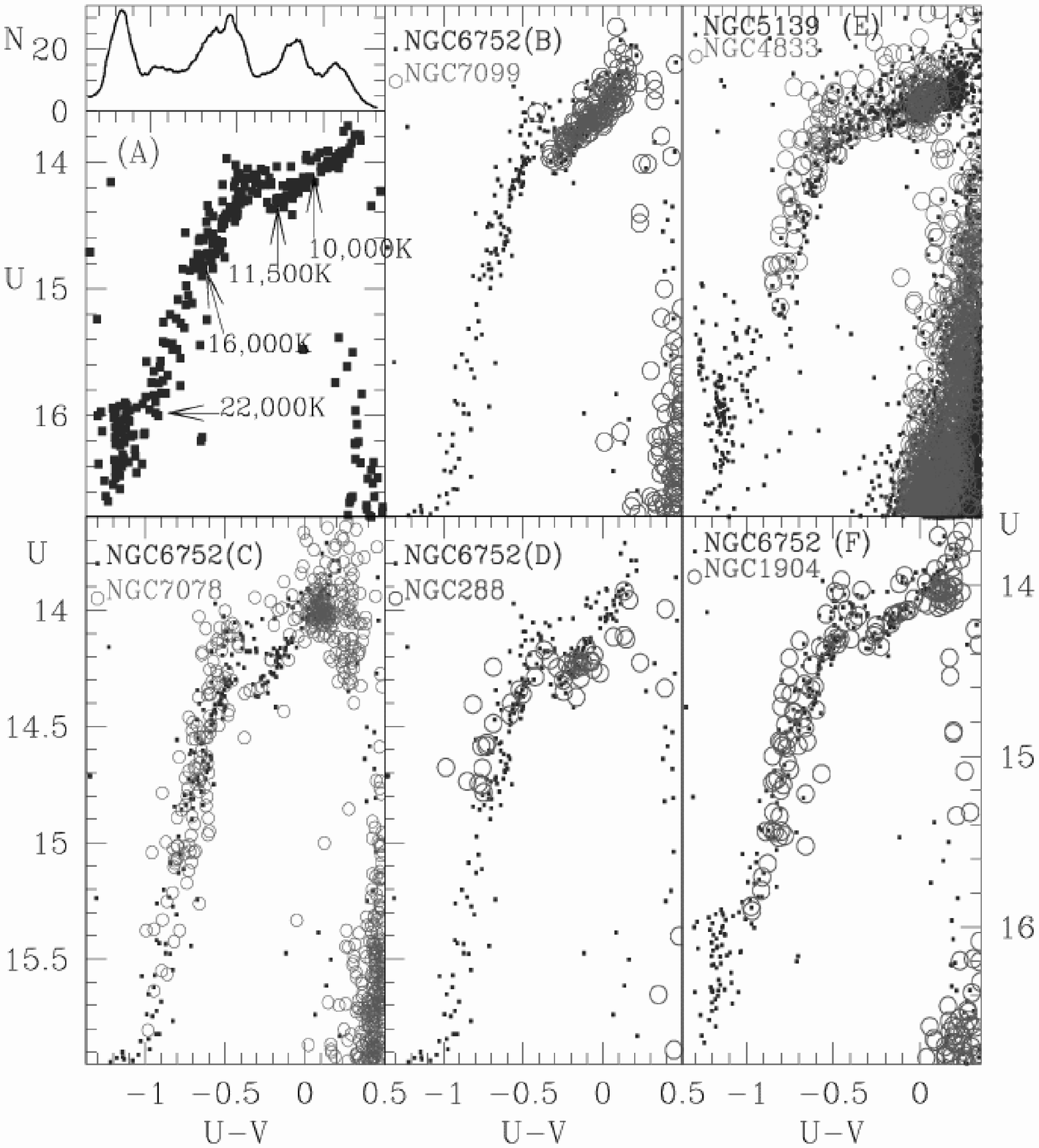}
\caption{The segmented nature of the blue HB. Panel (A) shows a zoom on 
the    HB  of NGC6752  along with     its color distribution   and the
identification of specific temperatures.   Panels (B) through (F) show
CMDs superpositions of    NGC7078,  NGC7099, NGC288   and NGC1904   on
NGC6752,  except  for   panel   (E) where   we overplot   NGC4833   on
$\omega$Cen.}
\label{f_twins}
\end{figure*}

\section{Summary}
\label{s_summary}

We have presented near-UV diagrams for a  dozen of clusters.  Our main
result is that the   previously reported discontinuity in NGC6752
(around $T_{\rm eff}\simeq 23,000$K)  is also present in other extreme
HB  clusters.  The  onset  of  this second  $U$-jump   in the
examined clusters seems also to coincide  at a  temperature of
 $\sim  21,000\pm 3,000$K.   Both  these facts strengthen  the idea
that the second $U$-jump is an indicator of a physical process, acting
in  all   extreme HB clusters,  that  has yet  to be   fully
understood.  We have clearly shown that the second $U$-jump population
is photometrically well disentangled from the  hotter blue hook stars,
and brought  evidence that the  two, chemically different, populations
can co-exist in objects like $\omega$Cen and M54.  In the flash mixing
scenario, early helium flashers are expected to pile  up at the end of
the extreme  HB, and this is  exactly what  the second jump population
seems to   show.  Hence, we suggest that    the second jump population
contains a significant number of early helium flashers.
%
%
%
%

%
\subsection{Final remarks and future work}

One of  the main difficulties in  studying the HB star distribution is
providing  a   satisfactory,   quantitative,    description  of    its
morphology. To this aim, many HB morphology parameters were introduced
(Fusi Pecci  et  al.\ \cite{fusi93},  Buonanno et  al.\ \cite{buon97},
Catelan et    al.\ \cite{cata98}, Piotto   et  al.\  \cite{piot99} and
references  therein) to describe the    distribution  of HB stars   in
color/temperature, measure  their maximum extension, and  reveal peaks
and gaps.  It remains however, that the  majority of these studies was
based on $BV$ CMDs, clearly not an ideal plane for detailed studies of
the HB (Ferraro et al.\ \cite{ferr98}).

In light of  the growing number of  $UV$ CMDs,  and recent photometric
findings  (the  G99  and second   $U$-jumps, and blue  hook stars)  we
revisit the  HB  morphology in  the $UV$ plane.   {\em  In  brief, the
analysis of  the HBs in 7 clusters  lead us to suggest  that the HB in
$UV$ CMDs can be envisaged as the sum of  discrete segments}.  This is
not a  new idea  (Buonanno  et al.\  \cite{buon85}).  However  the new
observations  in  the $UV$  plane and the  larger  sample  of HB stars
unveil a number of discrete branches in the HB previously undisclosed.
To better explain  this fact we rely  on Fig.~\ref{f_twins}, showing a
combination of $UV$ CMDs.  A more  detailed and complete analysis will
be subject  of   a future  work.   Figure~\ref{f_twins} allows  us  to
suggest the following:

\begin{enumerate}
\item The discontinuities (gaps or jumps) at $T_{\rm eff}\sim10,000$K 
and $11,500$K  occur  in {\em all}    blue HB   clusters,  in
particular, the G99   jump {\em coincides} with  the  endpoint of many
blue HB clusters and dwarf galaxies; 
\item In addition, there are two other {\em main} discontinuities at 
$T_{\rm eff}\sim16,000$K and  $21,000$K,  marking the endpoints of
HB reaching these temperature.
\end{enumerate}

Point (1) basically confirms the  already known ubiquitous nature
of the G99 jump,  and  suggests  a similar ubiquity  of the gap
at $T_{\rm eff}\sim10,000$K ({\it the gap at $B-V$  about zero} in the
Caloi \cite{caloi99} terminology).  
Similarly, Ferraro et al.\  (\cite{ferr98}) also suggested
that the gaps at $T_{\rm eff}\sim10,000$K  and $11,500$K (G0 and G1 in
their terminology)  occur in  {\em  many but not all  clusters}. 
Hence, the novelty in point (1) lies in suggesting the ubiquity of the
gap at $T_{\rm eff}\sim10,000$K, and this mainly relies on panels (B),
(C)  and  (D) of Fig.~\ref{f_twins}.    The three panels propose three
ways in which  clusters populate the blue  HB (i.e., the  HB extending
from the bluest boundary of the  RR instability strip  to the onset of
the G99 jump) around the $T_{\rm eff}\sim10,000$K gap.

The {\it first case}  (panel A overplotting NGC7099 on NGC6752)
is the most frequent, that is, a more or  less uniform distribution on
the two parts surrounding the gap at $T_{\rm eff}\sim10,000$K.
Relying  on the   optical CMDs of   the  HST snapshot (Piotto   et al.
\cite{piot02}), and   our $UV$ diagrams, we   count 34 clusters  with a
metallicity range between  [Fe/H]$=-1.27$ (NGC5904) and [Fe/H]$=-2.29$
(NGC5053)  which uniformly  populate {\em  only}  this part of  the HB
(i.e.  having no red HB   stars and no  HB  stars hotter  than the G99
jump).   Most  of these clusters   (e.g.   NGC7099) show  a
clear gap at $T_{\rm eff}\sim10,000$K.
On the other hand, we note that Local Group dwarf spheroidals are also
representative of a uniformly populated blue HB.  The diagrams of Ursa
Minor (Carrera et al.\ \cite{car02}) and Sculptor (Hurley-Keller et al.
\cite{hur99}) are perfect examples of blue HB  extending only in this
temperature  range,   possibly  showing   the   gap at    $T_{\rm
eff}\sim10,000$K gap.

The  {\it second and  third  case} (panels B  and  C) is when clusters
preferentially populate one side  of the $T_{\rm eff}\sim10,000$K
gap;  either  the hotter or   the cooler side.  NGC7078\footnote{Note
that the group of stars distributed on  the right side  of the blue HB
clump are probably variables stars (see Zheleznyak
\& Kravtosov \cite{zhel03}) caught at random phase.} and NGC5466 
(see  CMD in Buonanno et  al.\ \cite{buon85})  are  examples of blue HB
extending  from the RR    instability strip and  stopping  at  $T_{\rm
eff}\sim10,000$K; i.e. populating only the right side  of the gap.
On  the  other hand   NGC288 (panel  C)   shows an  opposite behavior;
populating  the  part between  $T_{\rm eff}\sim10,000$K and $11,500$K;
i.e.  populating the left side  of  the $T_{\rm eff}\sim10,000$K  gap.
Overall, the 3 proposed distribution modalities might explain previous
difficulties in ascertaining the ubiquity of  the gap at $\sim10,000$K
(see discussions in Ferraro et al.\ (1998) on  the {\em varying width}
of the $T_{\rm eff}\sim10,000$K gap from cluster to cluster).

We did not find
any cluster with HB extending beyond the G99 jump which do not possess
blue HB stars with  $T_{\rm eff}\le11,500$K.  In  other words, all the
clusters with hot HB stars  ($T_{\rm eff}\ge11,500$K) do have stars in
the region between the RR Lyrae instability strip and the G99 jump.

We do not have an explanation for such a complex behavior in this part
of the HB.    However, the presented   evidence implies that    the
discontinuities  at  $T_{\rm  eff}\sim10,000$K  and $11,500$K 
are  presumably present in  {\em all} clusters.   Whereas the G99 jump
can    be   easily discerned   in    UV  CMDS,   the   gap at  $T_{\rm
eff}\sim10,000$K might be  hampered by photometric  errors and post-HB
evolution. Moreover, given the high frequency of globular clusters and
dwarf galaxies populating only this specific range of  the blue HB (RR
instability  strip---onset  of the  G99 jump),  this group  of objects
might    represent the ``standard'' HB   morphology  in the metal-poor
regime.  As discussed   in Buonanno et  al.\ (\cite{buon85}),  one way
around the second parameter debate can be to isolate certain groups of
clusters (with   a substantial  similarity   in some   of their  basic
properties) and  then explore the effects of  any other difference the
clusters in the group  have.  Clusters/dwarf galaxies showing only the
blue  HB (between the RR instability  strip and  the  onset of the G99
jump) are most probably a separate group,  within which one can search
for ''similarities''.

{\it What is so special in the group of clusters (e.g. NGC7099) ending
their  blue HB  exactly   at the  onset of the    G99 jump at  $T_{\rm
eff}\sim11,500$}K ?  The occurrence of  the G99 jump was explained  as
the  aftermath of  radiative   levitation  that causes  a  substantial
increase  in  the metal content   of  the outermost layers.  Radiative
levitation  is  possible   after the  disappearance    of the envelope
convective layers  located across the  H and HeI ionization regions at
$T_{\rm eff}\sim10,000$K and $11,000$K respectively (Caloi
\cite{caloi99}, Sweigart  \cite{sweigart00}).  Hence, both the  gap at
$T_{\rm eff}\sim10,000$K and  the G99  jump  at $\sim 11,500$K can  be
attributed to atmospheric effects.
If we adopt  this explanation for  the two  discontinuities, and given
their omni-presence in  different environments, then systems with only
blue HBs ending just before the  onset of the G99 jump  can be seen as
clusters in which a ``standard'' mass  loss mechanism takes place. The
net product of this ``standard'' mass loss on the  red giant branch is
{\em always} HB with an envelope massive enough to possess an extended
convective region.

Similarly,    for   clusters  with     HB  extending  beyond   $T_{\rm
eff}\sim11,500$K,   we have identified  two  truncations points in the
blue tail:    one at  at  $T_{\rm  eff}\sim16,000$K  and  a  second at
$\sim21,000$K.


Panel (D) shows   that NGC288 ends at   $T_{\rm  eff}\sim16,000$K.  As
shown in the color distribution  of the HB  in NGC6752 (panel A), this
temperature corresponds to a marked  decrease in the stellar counts in
NGC6752. On the other hand, this is also the  temperature at which the
first of  the   NGC2808 gaps occurs   (Bedin  et al.\  \cite{bedin00}).
Hence,   besides the G99 jump,  $T_{\rm eff}\sim16,000$K seems to
mark    another endpoint.  Panels    (E) and (F)    indicate
another important   endpoint: the onset  of the second  $U$-jump.
Obviously one cannot rely on few stars to mark the end  of the  HB
blue  tail of NGC1904 and  NGC4833, however  the coincidence with the
onset of the second jump is rather tempting.
%
%
In  this regards,  it is of   great interest to  further investigate a
possible  relation between the   second jump population and the  early
helium flashers.    Just as a  working  hypothesis,  if  {\em all} the
second jump  stars were to  be early helium  flashers,  it would imply
that stars   hotter  than $T_{\rm  eff}\sim21,000$K  have  a different
physical origin (flash-mixing  scenario?)  with respect  to ``cooler''
HB stars (produced by standard mass-loss mechanism).  This possibility
would have  significant   implications on   our understanding of   the
complicated second parameter problem.

In conclusion,  the overall picture of  the HB in  $UV$ diagrams seems
rather segmented.   The endpoints which define these segments may
be acting  like markers, highlighting the  signature of  different
physical   processes  working in HB  stars. The  origin of the
$T_{\rm eff}\sim10,000$K and $\sim  11,500$K discontinuities seems  to
be related to  the   disappearance of the  convective  envelope layers
located  across   the H     and     HeI ionization  regions     (Caloi
\cite{caloi99}).   In this paper we  have shown that the discontinuity
at $T_{\rm  eff}\sim21,000$K can be  related to the presence  of early
helium flashers. For the   endpoint at $T_{\rm eff}\sim16,000$K we  do
not yet have an explanation.

\begin{acknowledgements}
We  warmly thank R.  Gratton for useful discussions, and  H.  Navasardyan for her
help in calibrating WFI data. We  acknowledge financial support of the
MIUR (PRIN 2001) and ASI.
\end{acknowledgements}

\begin{figure*}
\centering
\includegraphics[width=15cm,height=13.5cm]{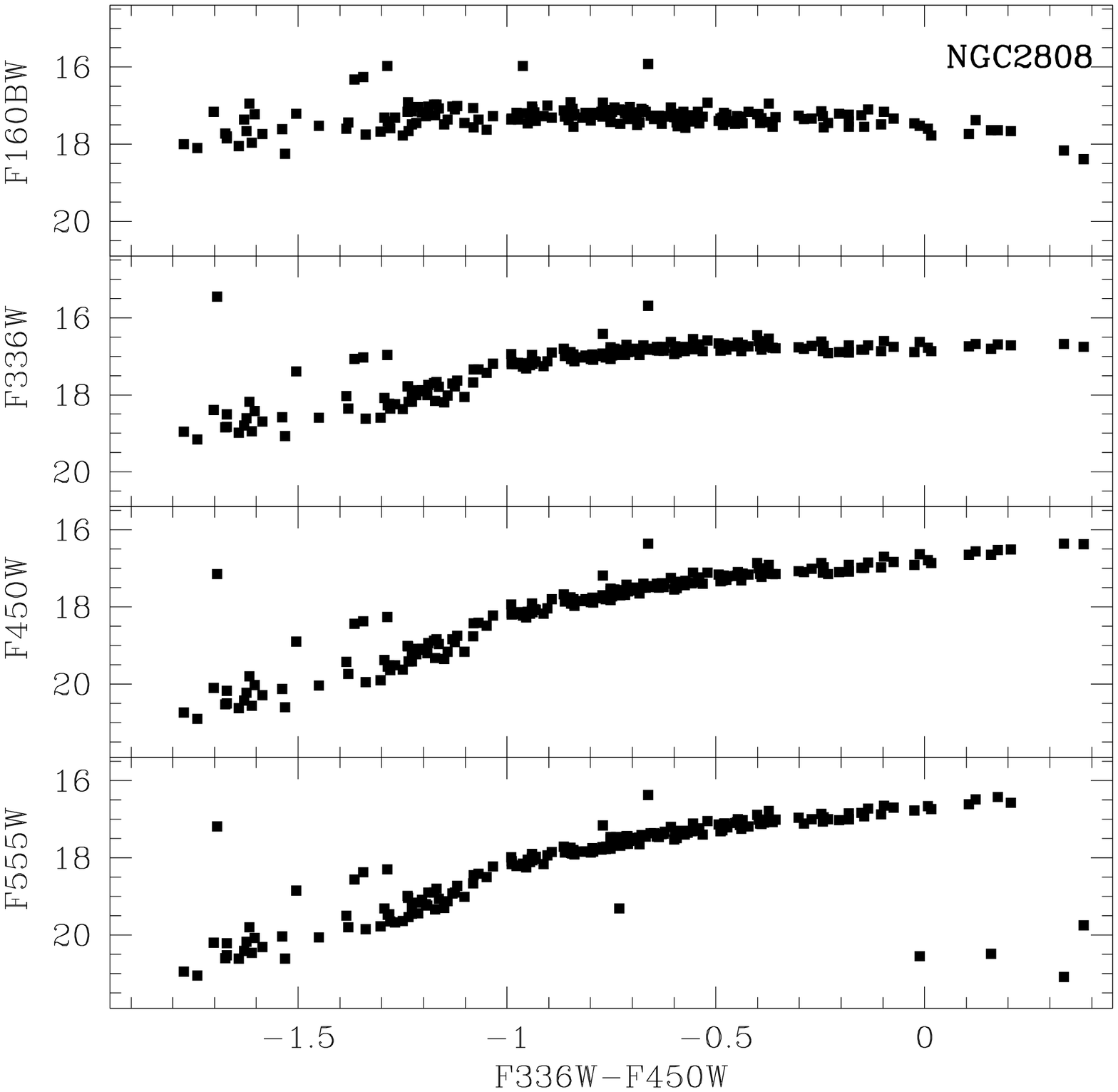}
\includegraphics[width=15cm,height=13.5cm]{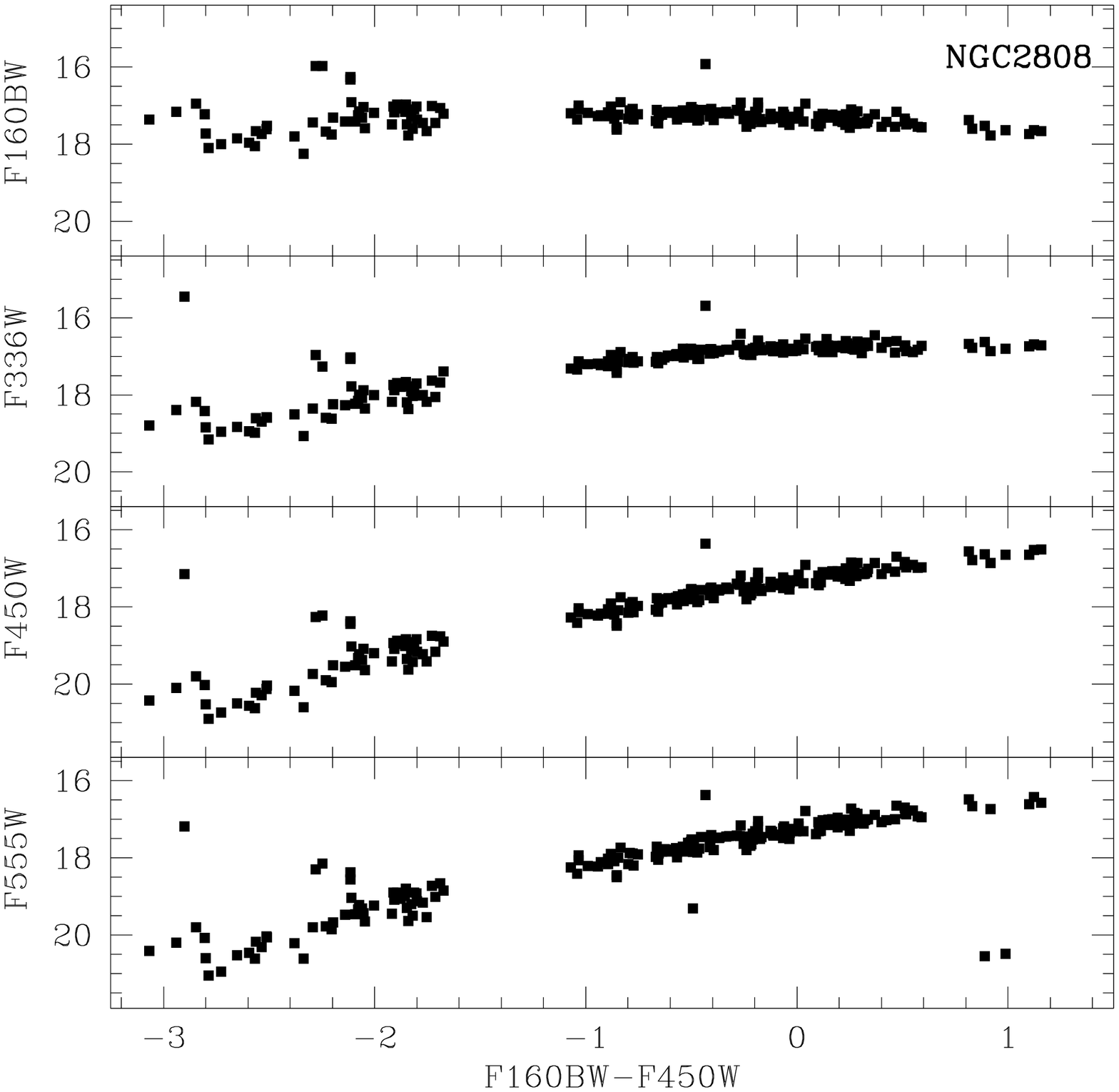}
\caption{{\bf NOT INCLUDED IN THE PAPER:} optical-far UV CMDs of NGC2808.
 Note  the different behavior  of gaps and  how the HB becomes almost
 horizontal.}
\label{f_FARUV}
\end{figure*}

\end{document}